## The Legacy of the Tevatron in the Area of Accelerator Science[1,2]

Stephen D. Holmes and Vladimir D. Shiltsev

Fermi National Accelerator Laboratory

P.O. Box 500

Batavia, IL  60510

e-mail: holmes@fnal.gov

shiltsev@fnal.gov

Abstract: For more than 25 years the Tevatron was the highest energy accelerator in the world, providing the first access to particle collisions beyond 1 TeV and achieving an ultimate performance a factor of four hundred beyond the original design goals. This article reviews the many formidable challenges that were overcome, and the knowledge gained, in building, operating, and improving the Tevatron over its lifetime. These challenges included: the first operations of an accelerator based on superconducting magnets, production of antiprotons in sufficient numbers to support a useable luminosity, management of beam-beam, intrabeam, and other collective effects, novel manipulations of the beam longitudinal phase space, and development and application of a wide variety of innovative technologies. These achievements established the legacy of the Tevatron as the progenitor of all subsequently constructed high energy hadron colliders.

1. Submitted to Annual Reviews of Nuclear and Particle Science

2. Work supported by the U.S. Department of Energy under contract number DE-AC02-07CH11359

# Contents

# 1. INTRODUCTION

The Tevatron (1) was one of the most productive particle accelerators ever built. For more than a quarter of a century, from startup in 1983 until it was surpassed by the Large Hadron Collider (LHC) in 2010, the Tevatron was the highest energy accelerator in the world. The Tevatron provided far ranging research opportunities based on both fixed target and colliding beam operations resulting in a multitude of discoveries and precision measurements that enhanced our understanding of nature. Among the more prominent scientific achievements were discovery of the top quark, discovery of the tau neutrino, discovery of direct CP violation in neutral kaon decays, observation and measurement of $B_s$ mixing, precision measurements of the W and Z masses and widths, observation and measurements of di-boson production, and the first hints of the Higgs boson (2, 3). All of these measurements coincided with, and contributed greatly to, the establishment of the Standard Model of particles and forces.

The Tevatron was a revolutionary machine from the perspective of accelerator physics and accelerator technologies (4). The Tevatron was the first synchrotron ever constructed utilizing superconducting magnets – it was the "existence proof" that served as the progenitor of all high energy colliders based on superconducting technologies that followed: the electron-proton collider HERA at the DESY laboratory in Germany (5), the Relativistic Heavy Ion Collider (RHIC) at Brookhaven National Laboratory in the U.S. (6), and the LHC spanning the France/Switzerland border at the CERN laboratory (7).

The idea of the Tevatron was conceived by Fermilab's founding Director, Robert R. Wilson, during the earliest stages of development of the laboratory (8). The original underground enclosure, housing the 500 GeV Main Ring synchrotron, was designed to accommodate the addition of a 1000 GeV ring based on superconducting magnets. The original name of this accelerator, the "Energy Doubler/Saver", reflected the joint mission of doubling the energy of the Fermilab accelerator complex while reducing power consumption. The development of superconducting magnets was begun in earnest following initial operations of the Main Ring in 1972, and culminated in the successful demonstration of accelerator quality magnets near the end of the decade. Construction of the Tevatron was authorized by the U.S. Department of Energy in 1979 and operations were initiated in 1983 (Figure 1). The initial beam energy of 512 GeV, raised to 800 GeV in 1984 and eventually to 980 GeV, established the Tevatron as the highest energy particle accelerator in the world.

While the Tevatron was initially operated in support of the Fermilab fixed target program, consideration of a colliding beam facility predated the final development of the superconducting magnets. In 1976 C. Rubbia and collaborators proposed the construction of a proton-antiproton collider at either CERN or Fermilab (9). The motivation for this proposal was the search for the W and Z bosons. In 1978 CERN initiated construction of the first proton-antiproton collider based on the SPS machine, utilizing the stochastic cooling technique developed by S. van der Meer and colleagues (10). The CERN collider commenced operations in 1981 at an energy of 630 GeV (center-of-mass), leading to the discovery of the W and Z bosons in 1983. Also in 1978 the new Fermilab Director, Leon Lederman, established the goal of achieving proton-antiproton collisions in the Tevatron. It was recognized that while the Tevatron could not beat the CERN machine to the W and Z, ultimately the higher energy would win out. First proton-antiproton collisions were achieved in the Tevatron in 1985, with the first extended run for data collection at the CDF (Collider Detector at Fermilab) experiment initiated in 1987. In 1992 a second collider detector, D0, was brought online. The Tevatron continued operations through September 30, 2011.

This article focuses on the achievements and legacy of the Tevatron in accelerator science and technology, with a focus on colliding beams (see also 11, 12, 13). The initially established performance goal for the Tevatron was a luminosity of $1 \times 10^{30}$ cm$^{-2}$sec$^{-1}$ at a center-of-mass energy of 1.8 TeV. When operations ceased a luminosity of $4.3 \times 10^{32}$ cm$^{-2}$sec$^{-1}$ at 1.96 TeV had been achieved. Many formidable challenges, arising both from technology and fundamental accelerator physics limitations, were overcome both to achieve the initial goal and the subsequent factor of four hundred improvement. Beyond the development of superconducting magnets these included: the production, cooling, and storage of antiprotons based on a variety of technologies; operations of an antiproton storage ring constructed of permanent magnets; dynamic lattice control; separated orbit operations; phase space manipulations of beams with radiofrequency systems; and control of a variety of collective effects including beam-beam interactions, intrabeam scattering, and coupled bunch instabilities. We review these challenges and look to the application of this knowledge to the continued development of energy frontier colliders.

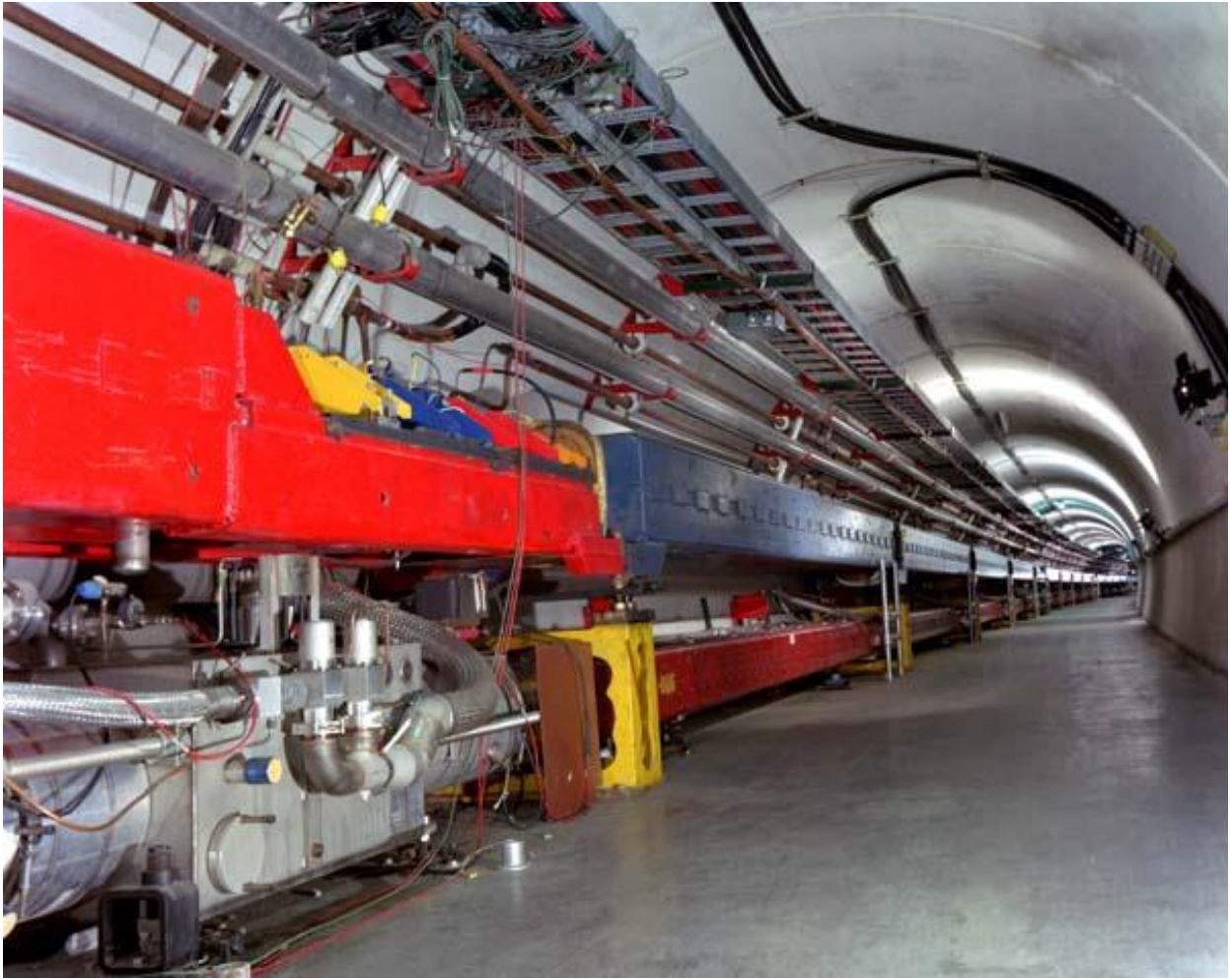

Figure 1: The normal-conducting Main Ring synchrotron (upper) and the Tevatron (lower) residing within their 6.3 km enclosure. During the initial stages of the Tevatron operations the Main Ring served as an injector into the Tevatron and as the source of protons for antiproton production. In the latter half of the Tevatron era this role was assumed by the Main Injector, and the Main Ring was decommissioned.

## 2. OVERVIEW OF THE TEVATRON COMPLEX AND PERFORMANCE STRATEGY

The establishment of proton-antiproton collisions within the Tevatron required the coordinated interaction of seven distinct accelerators and storage rings within the Fermilab complex. In this chapter we describe the operational roles of each of these accelerators in generating proton-antiproton collisions and the underlying strategy for maximizing delivered luminosity within the Collider as it evolved over two decades.

### 2.1 Summary of Operations of the Tevatron Complex

The layout of the Fermilab accelerator complex as it existed when the Tevatron ceased operations at the end of September, 2011 is shown in Figure 2. At that time the complex was supporting collider operations in the Tevatron in parallel with long baseline (NuMI) and short baseline (MiniBoone) neutrino programs. The accelerators in the figure are shown to scale with the radius of the Tevatron set at 1.0 km. Functionally the accelerators within the complex served three purposes: 1)production and delivery of protons to the Tevatron; 2)production and delivery of antiprotons to the Tevatron; 3)generation of high energy collisions of protons and antiprotons in the Tevatron.  Among the accelerator shown in Figure 2 all remain in operation today, although with somewhat different roles, with the exception of the Tevatron.

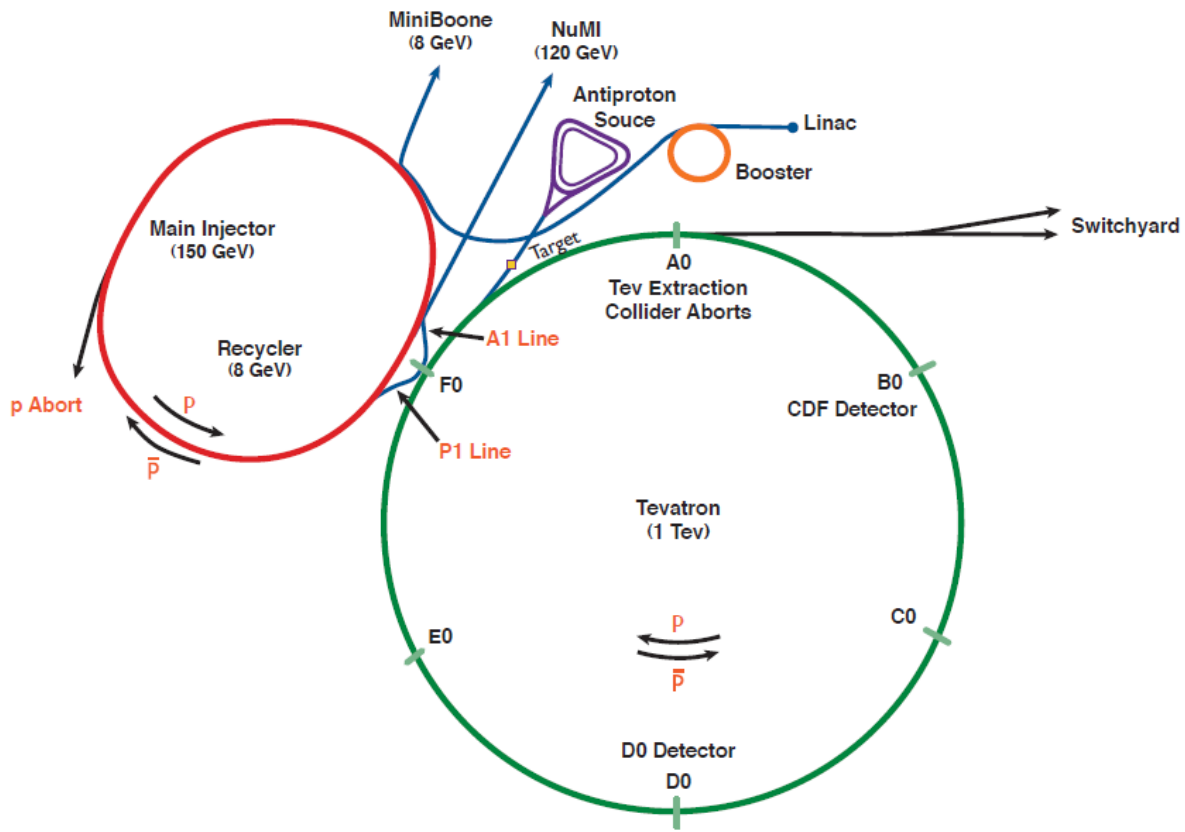

Figure 2: Layout of the Fermilab accelerator complex in 2011.

### 2.1.1 The Proton Source

The Proton Source consists of a 750 kV Cockcroft-Walton pre-accelerator, a 400 MeV[3] linear accelerator, an 8 GeV Booster synchrotron, and the 150 GeV Main Injector synchrotron. The linac operates with a (H⁻) beam current of 35 mA with a pulse length of 40 μsec. The linac energy was raised from 200 MeV to 400 MeV in 1992.

---

[3] A note on units: All quantities expressed in eV represent kinetic energy; quantities expressed in eV/c represent momentum.

The 8 GeV Booster is a rapid-cycling synchrotron that accepts beam from the linac and accelerates it to 8.9 GeV/c with a 0.067 sec acceleration cycle. The Booster utilizes a multi-turn (typically 10-15 turns) injection that passes the incoming H$^-$ ions through a carbon stripping foil, thereby merging the incoming and circulating proton beams onto a common orbit. The Booster typically delivers $(4.5\text{-}5.0)\times10^{12}$ protons per pulse to the Main Injector at 8 GeV. Booster performance improved dramatically following the increase in injection energy to 400 MeV.

The Main Injector (MI) synchrotron (14) accepts protons from the Booster and accelerates them to 120 or 150 GeV. The Main Injector is seven times the circumference of the Booster, and hence filling requires repeated Booster injections. The Main Injector provided beams for multiple purposes including delivery of protons and antiprotons to the Tevatron Collider, and delivery of protons to both the antiproton and neutrino production targets. Typical performance of the Main Injector at the end of Collider operations was $8.0\times10^{12}$ protons delivered to the antiproton production target and $3.5\times10^{13}$ protons delivered to the neutrino production target at 120 GeV every 2.2 seconds. For direct support of the Tevatron the Main Injector delivered 36 bunches containing about $3\times10^{11}$ protons each at an energy of 150 GeV. The Main Injector initiated operations in 1999, replacing the original Main Ring synchrotron.

2.1.2 The Antiproton Source

The Antiproton Source generated, collected, cooled, and stored the antiprotons required to support the Tevatron Collider. The Antiproton Source consisted of a target station and a set of 8.9 GeV/c storage rings: the Debuncher, the Accumulator, and the Recycler (15, 16). At the end of Tevatron operations the Antiproton Source was capable of producing up to $2.5 \times 10^{11}$ antiprotons/hour, with a total storage capacity of up to $6.0 \times 10^{12}$ antiprotons.

The Debuncher and Accumulator shared a common enclosure. Negatively charged particles collected off the production target were transported via a large aperture beamline into the Debuncher. Upon entering the Debuncher the individual bunches were rotated in phase space to reduce the momentum spread (by roughly a factor of ten) with an accompanying increase in the bunch length – hence the name "Debuncher". The Debuncher operated on the same cycle as the MI, delivering about $2 \times 10^8$ antiprotons to the Accumulator every 2.2 seconds.

The Accumulator accepted antiprotons from the Debuncher and stacked them in momentum space utilizing stochastic cooling. The Accumulator was capable of accumulating up to $\sim 1.5 \times 10^{12}$ antiprotons before a decreasing stacking rate led to diminishing returns. Following construction of the Recycler Accumulator stacks were limited to about $3 \times 10^{11}$ antiprotons with transfers for long term storage in the Recycler every ~30 minutes.

The Recycler shares the Main Injector enclosure and is the only large scale storage ring in the world constructed of permanent magnets. With both stochastic and electron cooling capabilities, and at seven times the circumference, the Recycler was able to accommodate significantly larger numbers of antiproton than could the Accumulator – up to $6 \times 10^{12}$ antiprotons, although in routine operations $(350\text{-}450) \times 10^{10}$ antiprotons were more typical. The Recycler supplied antiprotons to the Main Injector for acceleration to 150 GeV and injection into the Tevatron.

### 2.1.3 The Tevatron

The Tevatron was a 6.3 km circumference superconducting synchrotron that accepted proton and antiproton beams from the Main Injector at 150 GeV and accelerated these beams to 980 GeV (4). Collisions were provided at two interaction points, designated B0 and D0, accommodating the CDF and D0 experimental detectors. Each beam contained 36 bunches arranged in three trains of twelve bunches with an intra-train bunch spacing of 396 nsec. Each bunch typically contained $3\times10^{11}$ protons or $1\times10^{11}$ antiprotons. Since both beams circulated within a common vacuum beampipe an electrostatic separator system was required to ensure that the beams collided in only two locations around the ring.

### 2.1.4 Low-beta Squeeze

The introduction of a "low-beta squeeze" was an essential and innovative feature of Tevatron Collider operations. Because the maximum beta functions in the interaction regions were inversely related to the beta function at the interaction point ($\beta$*), and because the beam size was largest at injection, the optics generated by a $\beta$* <1 meter could not be utilized during Tevatron injection. The solution was to utilize an "injection lattice" optics with a $\beta$* of 1.7 m and then, following beam acceleration, transition to a $\beta$* of 0.3m. This operation had to be completed without generating beam loss (which could quench the magnets) or emittance dilution. All quadrupoles within the low beta region were adjusted in a series of twenty-five steps to provide a smooth transition between the two sets of optics. This process took about 2 minutes.

## 2.1.5 Operational Cycles

The Tevatron Collider operated on an approximately 24 hour cycle. A cycle was initiated with a "shot", which refers to the delivery of protons and antiprotons to the Tevatron followed by acceleration to 980 GeV and the initiation of collisions. A typical shot proceeded as follows: Thirty-six proton bunches each were injected onto the Tevatron central orbit. The electrostatic separators were then powered to move the protons onto a helical orbit. Antiproton bunches were injected four at a time into gaps between the three proton bunch trains. An rf manipulation ("cogging") was utilized to move antiprotons out of the injection gap in order to provide space for subsequent injections. Once beam loading was complete, the beams were accelerated to 980 GeV and the Tevatron optics were adjusted to the collision configuration (low-beta squeeze). Finally the two beams were cogged so that they collided at CDF and D0 and halo was reduced using the collimator system. At this point collisions were produced for an extended period of time (typically ~20 hours), called a "store". While the store was in progress, a new complement of antiprotons was prepared for the subsequent shot.

## 2.1.6 The Sequencer

The coordinated operations of the accelerators making up the Tevatron complex were dependent on a computer meta-program called the Collider Beam Sequencer. The Sequencer synchronized the application programs, and simulated the activities of accelerator operators, required to control each accelerator during a shot or a store. The Sequencer issued high level control commands, adjusted accelerator timelines, launched programs and scripts for complex operations, and listened to and set "state devices" to provide the required synchronization (17). There were separate sequencer instances for the Tevatron, Main Injector, Accumulator, Debuncher,

Recycler, etc. and for particle transfers from one machine to another. For example, the aggregate commands of the Tevatron sequencer included 14 states: "proton injection porch", "proton injection tuneup", "reverse injection", "inject protons", "antiproton injection porch", "inject antiprotons", "cogging", "before ramp", "acceleration", "flattop", "squeeze", "initiate collisions", "remove halo" and "HEP". The Sequencer automation was critical in maximizing store hours, minimizing shot set up time, and reducing time needed for transitioning between operational modes.

In parallel the Sequencer recorded vast amounts of accelerator data crucial to evaluating and improving machine performance, and diagnosing failures. The readings and settings of accelerator devices were obtained via Fermilab's Accelerator Controls Network (ACNET) control system. Device data could be plotted live at up to 1440 Hz or logged at various fixed rates or periods, e.g., 15 Hz or 1 minute, or on a specific event, e.g. when the energy ramp is complete or at each antiproton injection. Data were stored in circular buffers with a wrap-around time dependent on the number of devices and their logged rate. Logged data up to a 1 Hz maximum rate were also copied to a "backup" logger for long-term storage and processed by Shot Data Analysis (SDA) software (18). The SDA automatically generated summary reports and tables for each store. These data were readily accessible by various means and allowed for convenient analysis of the accelerator complex on a shot-by-shot basis.

## 2.2 Performance Limitations and the 20-year Strategy

The luminosity achievable in the Tevatron can be written as:

$$L = \frac{f_0 n_b N_p N_a}{2\pi \left(\sigma_p^2 + \sigma_a^2\right)} H\left(\frac{\sigma_z}{\beta^*}\right) = \frac{\gamma f_0 (N_p / \varepsilon_{pn})(n_b N_a)}{2\pi \beta^* \left(1 + \varepsilon_{an} / \varepsilon_{pn}\right)} H\left(\frac{\sigma_z}{\beta^*}\right) \qquad (1)$$

Here $f_0$ is the Tevatron revolution frequency, $n_b$ is the number of bunches in each beam, $N_{p,a}$ is the number of protons and antiprotons in a bunch, $\sigma_{p,a}$ is the rms transverse beam size at the interaction point (equal in the horizontal and vertical planes), $\gamma$ is the relativistic beam energy, $\varepsilon_{p,a\ n}$ is the rms normalized transverse beam emittance, $\beta^*$ is the optical beta function at the interaction point, $\sigma_z$ is the rms bunch length and $H$ is a geometrical form factor ($<1$). The luminosity formula highlights the most significant factors determining luminosity performance:

- *The proton beam density, $N_p/\varepsilon_{pn}$, is directly proportional to the beam-beam tune shift experienced by the antiprotons for each head-on encounter with the protons.*

- *The total number of antiprotons, $n_b N_a$, depends on the antiproton accumulation rate and the ability to cool and store a suitably large number of antiprotons.*

- *The beta function at the interaction point, $\beta^*$, is inversely proportional to the luminosity, subject to limitations from the "hour-glass" factor, $H(\sigma_z/\beta^*)$, which approaches one as $\sigma_z/\beta^*$ becomes much less than one.*

### 2.2.1 Strategy

A long term strategy for improving performance beyond the original design goals was developed in the late 1980's (19) and was executed in two stages with the corresponding operational periods named "Run I" and "Run II". Run I was established with a goal of achieving a luminosity of $1\times10^{31}$ cm$^{-2}$sec$^{-1}$, a factor of ten beyond the original design. The primary elements of the Run I strategy were:

- Electrostatic separators: Twenty-two electrostatic separators were installed to enable operations with protons and antiprotons traveling on separated helical orbits. The separators mitigated the beam-beam effect and allowed an increase in the number of bunches as the luminosity rose.

- Low-beta magnet systems: Two sets of high performance quadrupoles providing matched optics were developed and installed at B0 and D0.

- 400 MeV linac upgrade: The linac energy was increased to 400 MeV to reduce space-charge effects at injection into the Booster, thereby providing higher beam intensity.

- Antiproton Source improvements: Improvements to stochastic cooling and target systems were made to accommodate the higher antiproton flux generated by continuously increasing numbers of protons on the production target.

These improvements supported an increase in luminosity to $1.6 \times 10^{31}$ cm$^{-2}$sec$^{-1}$ over the period of the 1990's. A subsequent set of improvements was implemented in the late 1990's in support of Collider Run II, conducted within the 2000's (20):

- Main Injector: The Main Injector replaced the original Main Ring as a larger aperture, faster cycling synchrotron (14). The goal was to increase the antiproton accumulation rate, accompanied by the ability to obtain good transmission from the Antiproton Source to the Tevatron from large antiproton stacks.

- Recycler: The Recycler was designed to provide storage for very large numbers of antiprotons – well beyond the capabilities of the Accumulator.

The Main Injector and Recycler enabled performance roughly a factor of twenty beyond Run I, and a factor of 300-400 beyond the original Tevatron design. The comparison between the original design goals and the achieved performance at the end of Collider operations is given in Table 1.

| | Tevatron Design | Tevatron 2011 Actual | |
|---|---|---|---|
| Energy (center-of-mass) | 1800 | 1960 | GeV |
| Revolution frequency, $f_0$ | 48 | 48 | kHz |
| Protons/bunch, $N_p$ | $6 \times 10^{10}$ | $29 \times 10^{10}$ | |
| Antiprotons/bunch, $N_a$ | $6 \times 10^{10}$ | $8 \times 10^{10}$ | |
| Number of bunches, $n_b$ | 3 | 36 | |
| Total Antiprotons, $n_b N_a$ | $18 \times 10^{10}$ | $290 \times 10^{10}$ | |
| Proton emittance (rms, normalized), $\varepsilon_{pn}$ | 3.3 | 3 | $\pi$ mm-mrad |
| Antiproton emittance (rms, normalized), $\varepsilon_{an}$ | 3.3 | 1.5 | $\pi$ mm-mrad |
| IP beta-function, $\beta*$ | 100 | 28 | cm |
| Luminosity | $1 \times 10^{30}$ | $340 \times 10^{30}$ | cm$^{-2}$sec$^{-1}$ |

Table 1: Achieved performance parameters at the end of Tevatron Collider operations in comparison to original design values. All listed performance parameters represent typical values at the beginning of a store.

# 3. ADVANCES IN ACCELERATOR SCIENCE AND TECHNOLOGIES

The successful construction and initial operations of the Tevatron required the development of a large number of technically advanced systems, ranging from superconducting magnets to stochastic cooling systems. As performance evolved to levels substantially beyond initial design goals new limitations were uncovered, originating primarily from beam dynamics phenomena. In this chapter we describe the underlying technical and accelerator science bases for the Tevatron, and their evolution as knowledge was gained and means to break through limitations were identified and implemented.

## 3.1 Magnets

The Tevatron was the first high energy accelerator to utilize superconducting magnet technology on a large scale. Twenty years after the development of superconducting magnets, a different path was taken to meet the needs for enhanced antiproton storage capacity – the Recycler became the first storage ring based primarily on permanent magnets to be brought into operations.

### 3.1.1 Superconducting Magnets

The Tevatron was constructed of 774 dipole and 216 quadrupole magnets. These magnets utilized NbTi as the superconductor, operating at 4.2 K and developed through an effective industrial partnership. The dipole field corresponding to 1 TeV was 4.4 T and the quadrupole gradient was 70 T/m. The development of these magnets was initiated in the mid-1970's.  In the late 1980's high performance quadrupoles were developed for improved optics at the B0 and D0 interactions regions. Twenty-four of these "low beta quadrupoles" were installed in the Tevatron in 1991, with operational gradients up to 140 T/m.

The technical issues associated with the construction and operation of superconducting magnets for accelerators are described in detail by Tollestrup and co-authors (21,22), while the development of the low beta quadrupoles is described in (23,24,25). Here we describe some of the characteristics of these magnets that manifested themselves in terms of operational performance limitations.

Persistent Currents and Snap-back Phenomena

Eddy currents produced in the superconducting magnet coils as a result of ramping of the fields tend to persist, subject to limitations imposed by the critical current within the superconductor. The net result of these "persistent currents" is the generation of fields that depend on the magnetic history, the geometry of the coil, and the superconducting filament size. In general the most important field perturbation is a sextupole field, which can be large enough to impact accelerator performance substantially. A number of features associated with persistent currents were first observed in the Tevatron, and mitigation of these features has been incorporated into the design of all superconducting accelerators since.

The strength of persistent currents is generally characterized by an anomalous sextupole component within the dipole magnet. The effect can be seen in the hysteresis loop corresponding to a complete cycle of a Tevatron dipole measure on a test stand in which injection, acceleration, store, deceleration, and reset are simulated in a controlled manner – see Figure 3. Two features displayed in the figure were first observed and mitigated in the Tevatron: 1) the gradual decay of persistent currents when the magnets were held at a fixed current during the extended injection period (26); and 2) the rapid reversion, over a few seconds, of these persistent currents to the hysteresis curve at the start of acceleration (27). In terms of chromaticity the magnitude of these effects was of order 50 units – a major perturbation to the optical characteristics of the Tevatron. Left uncorrected these lead to both beam loss and emittance dilution (via beam instabilities).  In practice beam loss could be sufficient to quench the Tevatron magnets.

A number of strategies were employed to mitigate these effects. The persistent current drift at injection was recognized and dealt with first (28). The sextupole component of the magnetic field was parameterized with a logarithmic time dependence and automated adjustments were made to the correction sextupoles to compensate. Prior to injection it was found to be necessary to send the Tevatron through six acceleration/deceleration cycles (without beam) to establish reproducible conditions. Ultimately it was also necessary to characterize the snap-back phenomena and again to adjust the correction sextupoles during the first few seconds of acceleration (29).

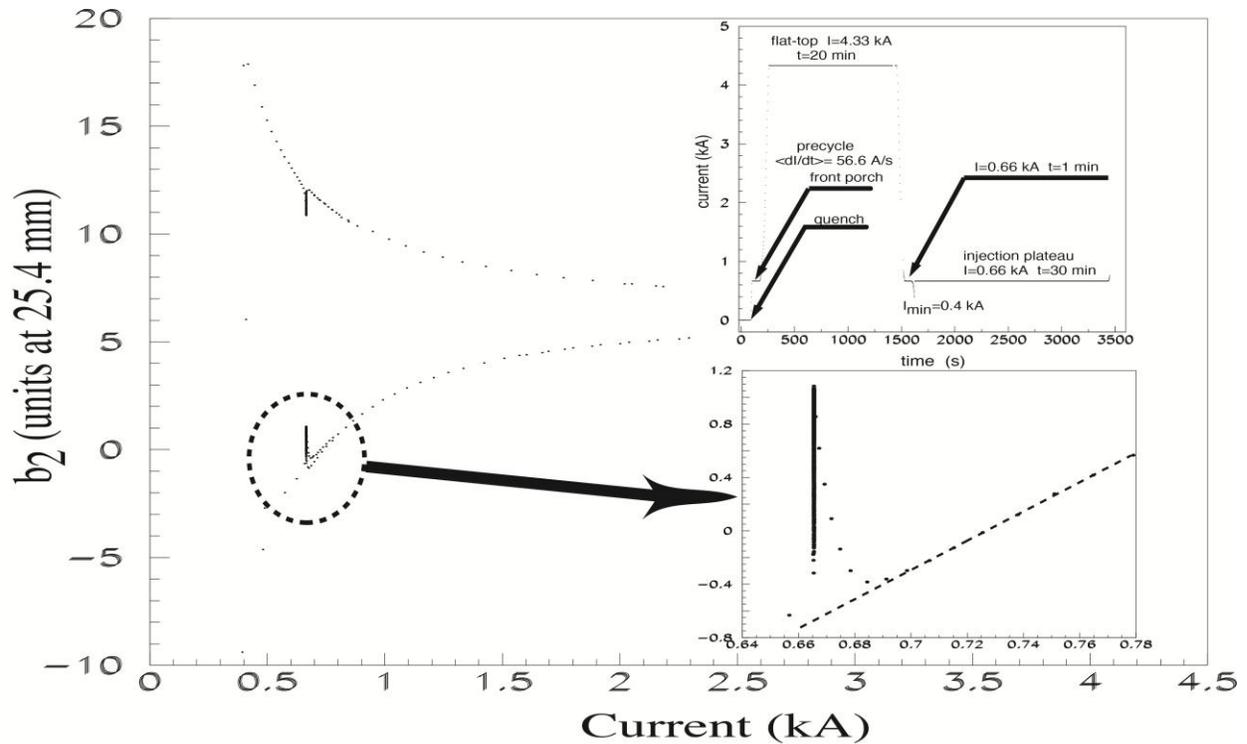

Figure 3: Sextupole component (b2) of the Tevatron dipole magnet as a function of current. The magnet current waveform corresponding to the hysteresis curve is shown in the upper right. An expansion of the curve corresponding to the injection energy (150 GeV), is shown in the lower right. Both the decay of the persistent current with time and the snapback are visible.

Skew Quadrupole Field Components Introduced via Mechanical Creep

A more subtle, and unanticipated, phenomenon was the development over two decades of a significant skew quadrupole component in the Tevatron superconducting dipoles. The effect manifested itself through very strong coupling between the horizontal and vertical beam motion, characterized by a minimum tune split, $Q_H - Q_V$, of about 0.3. The coupling source was observed to be distributed, not localized (30,31). This strong coupling made control of the beam difficult and degraded luminosity performance.

A cross section of the Tevatron dipole magnet is shown in Figure 4. This magnet was constructed from a "cold-mass" suspended within a warm iron yoke. The warm iron contributes significantly to the magnetic field. The suspension system consisted of two fixed bolts below and two spring loaded bolts above the cold mass with fiberglass (G11) spacers at the warm-cold interfaces to minimize the heat leak. This design allowed shimming of the spacers during the original cold test stand measurements to produce a uniform field. It was discovered that over two decades mechanical creep in the G11 spacers had produced a systematic offset of the cold mass relative to the yoke. The average offset, about 0.15 mm, was sufficient to explain the observed large coupling in the Tevatron. In response all Tevatron dipoles were reshimmed, in situ. With a total of 18 shims per magnet this was a massive undertaking and was completed over three accelerator shutdowns during the period 2003-2006. Performance improved significantly following the first shutdown and coupling from this source was effectively eliminated at the end of the third session (32).

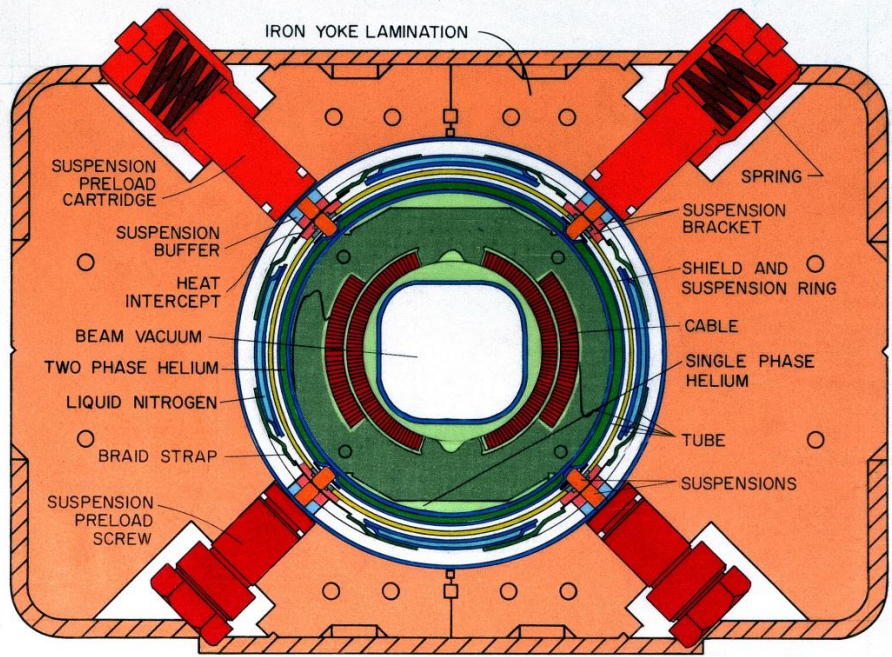

Figure 4: Cross section of the Tevatron superconducting dipole magnet.

### 3.1.2 Permanent Magnets

Because of the low ratio of beam momentum to circumference, the Recycler was uniquely suited for utilization of permanent magnets (33,34). The advantage of permanent magnets was the low construction and operating costs – the Recycler may well be the lowest cost per GeV accelerator/storage ring ever constructed.

The Recycler lattice optics replicate those of the Main Injector. The arcs are based on combined function magnets including dipole, quadrupole, and sextupole fields. Permanent magnet quadrupoles are also utilized in the straight sections. In total 362 gradient magnets, 109 quadrupole magnets, and 13 specialty permanent magnets make up the Recycler. The required central dipole field in the combined function magnets is about 0.14 T. Tune and orbit adjustment are provided by electromagnetic trim dipoles and trim quadrupoles.

The magnets were constructed of precision shaped iron pole pieces backed by strontium ferrite bricks. The primary issues were: 1) achieving field uniformity sufficient to support stored beams; 2) stability against temperature variations; and 3) long term drift. Field uniformity relative to the nominal fields of a few parts in $10^4$ was achieved by measuring field components up to 12-pole on each magnet during production, and then calculating an end shim shape that would correct the quadrupole, skew quadrupole, sextupole, and octupole components accordingly. The calculations were fed to a computer controlled electric discharge machine (EDM) that produced the end shims within 24 hours. Nearly all magnets were adequately corrected on the first iteration.

Initial magnet measurements showed a strength variation with temperature of about -0.2% per $^o$C. A compensation scheme was implemented to allow stable operations within the enclosure environment. A nickel-iron alloy with a low Curie temperature (in the range 40-45 $^o$C) was deployed (longitudinally) between the ferrite bricks to shunt flux in a temperature dependent manner and cancel the natural temperature variation. With the inclusion of the compensator the magnet strength variation was reduced to about 0.004% per $^o$C.

Finally, the long-term stability of the fields was a concern. A prototype magnet was monitored during the entire year-long magnet production run and for several years thereafter. The magnet strength was found to decay with a logarithmic time dependence, losing about 0.1% of its initial strength over the first two years. The best long term measure of strength comes from monitoring the revolution frequency of the Recycler – we infer an additional 0.04% decline in strength over the following decade. No operational difficulties have been attributed to these changing fields which are readily accommodated via adjustment to the upstream and downstream accelerators.

## 3.2 Antiproton Production and Cooling

The utilization of antiprotons in the Tevatron created challenges in production, collection, and cooling to form suitably dense beams. Antiprotons were produced by directing 120 GeV protons onto a solid target (initially nickel, ultimately Inconel), and collecting negatively charged secondary particles at 8.9 GeV/c, i.e. approximately at rest in the beam-target center of mass where antiproton production is maximized. The collected antiprotons were stored and cooled in a series of rings, also operating at 8.9 GeV/c. The acceptance of the collection systems was $30\pi$ mm-mrad transversely and 4% in momentum space. The collected antiproton phase space density was many orders of magnitude less than that required to make usable luminosity. Hence "cooling" systems were required to increase the beam density.

Through the 1990's antiprotons were collected and stored in two rings, the Debuncher and the Accumulator, utilizing stochastic cooling to increase the antiproton phase space density. In the mid-2000's the Recycler came into operations, utilizing electron cooling for further phase space reduction. Stochastic and electron cooling offer complementary strengths and weakness – the cooling rate in a stochastic cooling system decreases inversely with the number of particles being acted upon in the system; the cooling rate in an electron cooling system is independent of the number of particles but decreases rapidly with increasing particle energy. The joint systems in the Recycler fully integrated the best aspects of both systems leading to a capability to accumulate and cool approximately 20 antiprotons for every one million protons striking the production target, at a production rate of about $2.5 \times 10^{11}$ antiprotons per hour.

The production and accumulation of antiprotons generally followed the scheme developed at CERN (35). Fermilab built on this scheme initially with two important innovations (15): 1) the development of Lithium lenses for antiproton collection immediately downstream of the production target; and 2) the introduction of a "Debuncher Ring" that reduced the momentum spread of the beam collected off the target prior to delivery to the stochastic cooling systems. Both of these features were subsequently adopted by CERN for the upgrades of their antiproton facility.

Lithium is the lowest Z conductor and hence provided an ideal material for a collection lens. The lens was formed as a 1 cm (radius) by 15 cm (length) cylinder. An axial current of 400 kA produced a gradient of about 900 T/m. This arrangement produced an azimuthal field which focused simultaneously in both transverse dimensions, thereby maximizing acceptance. The beam impinging on the production target was configured to have a very short bunch length (typically 1 nsec full width) which was retained by the antiprotons produced in the target. These antiprotons, with their accompanying large momentum spread (4%), were captured in the Debuncher where a longitudinal bunch rotation exchanged the short bunch length and large momentum spread for a long bunch length and small momentum spread (0.4%). These antiprotons were then passed onto the Accumulator for accumulation, storage, and cooling.

### 3.2.1 Stochastic Cooling

The accumulation and packaging into beams of antiprotons with useful densities was dependent on stochastic cooling, a technology developed and implemented at CERN (10). The basic idea of stochastic cooling is that a "pickup" is used to sense the centroid of a beam distribution and this information is then transmitted to a "kicker" that applies a correction characterized by a gain, $g$. The essential factors governing the performance of the cooling system are captured in the expression for the time evolution of the transverse beam emittance, $\varepsilon$, of an ensemble of $N$ particles:

$$\frac{1}{\varepsilon}\frac{d\varepsilon}{dt} = -\frac{W}{2N}[2g - g^2(M + U)] \qquad (2)$$

The basic cooling mechanism can be understood as follows: The bandwidth of the system, $W$, represents a resolving time ($\sim 1/W$). Because the bandwidth is finite particle positions are not sampled individually but in groups, and an individual particle is only cooled via subsequent applications of a correction if the particles accompanying it are randomized periodically. Noise in the system, $U$, represents a heating term. The mixing factor, $M$, can be thought of as the number of traversals of the pickup-kicker system that a beam sample undergoes before becoming completely randomized – $M$ is greater than or equal to 1. An examination of the cooling equation shows that there is an optimum gain and a corresponding optimum cooling rate:

$$\frac{1}{\tau_{cool}} = \frac{W}{2N}\frac{1}{M + U} \qquad (3)$$

One sees that the cooling rate goes down as the number of particles being cooled increases and that the critical parameters governing the cooling rate are the bandwidth, the mixing factor, and the signal to noise ratio in the amplifier. The noise contribution was minimized by operating the amplifiers at cryogenic temperatures – 80 K initially and 4 K by the end of Antiproton Source operations. Momentum cooling proceeds in a similar manner if one places the pickups in a region of high dispersion, generating a momentum-position correlation. Alternatively a filter cooling system can be utilized to provide a correction signal based on the revolution frequency of the particles. Both types of systems were utilized in the Antiproton Source at Fermilab.

The unique contributions of the Fermilab facility were the development of very high bandwidth (4-8 GHz) pickups and kickers based on planar arrays and slotted waveguides, the utilization of free space (laser-driven) links between pickup and kicker, and the ability to vary the mixing factor as the operational demands required (36). The development of novel pickup and kicker technologies is described in (37,38). These innovations were key to providing the necessary cooling system bandwidth to keep pace with the faster accumulation rates associated with continual increase in the number antiprotons collected off the production target, primarily through the continual increase in the proton flux onto the target. Ultimately the Antiproton Source featured stochastic cooling systems with a bandwidth covering 1 – 8 GHz (39).

### 3.2.2 Electron Cooling

The ability to accumulate sufficient antiprotons to support ever increasing performance requirements in the Tevatron was ultimately limited by the stochastic cooling systems in the Accumulator. To overcome these limitations the Recycler (16) was constructed and an aggressive development program to introduce electron cooling was initiated.

Cooling by electrons was first described and demonstrated by G. Budker and colleagues in Novosibirsk (40,41). Electron cooling has since been extended to the cooling of non-relativistic protons and ions in a variety of facilities (42). The electron cooling system introduced into the Recycler was unique in two respects (43): 1) the cooling was applied to a relativistic (antiproton) beam; and 2) the system utilized "non-magnetized" cooling. Cooling occurred in a 20 m section in which 4.3 MeV electrons mixed with circulating 8.9 GeV/c antiprotons. The electron beam was supplied by an electrostatic generator ("Pelletron") capable of sourcing 100 μA. A total electron beam current of 500 mA was sustained by returning 99.98% of the electron beam for deceleration through the accelerating column. Within the cooling section the electron beam was round with a radius of 3.5 mm, an angular divergence of <0.2 mrad, and an energy spread of 200 eV. The beam transport through the cooling channel utilized weak (100 G) solenoids.

The first observation of electron cooling of antiprotons in the Recycler came in 2005 (Figure 5). The system ultimately proved capable of providing a maximum cooling force of about 25 MeV/hour. In practice this allowed the Recycler to store antiproton beams containing up to $6 \times 10^{12}$ particles with a total longitudinal emittance of 60 eV-sec, and a transverse emittance of 0.5 mm-mrad (rms, normalized). The outstanding performance of electron cooling proved critical to the ultimate performance of the Tevatron – no luminosity above $10^{32}$ cm$^{-2}$s$^{-1}$ was ever achieved without electron cooling.

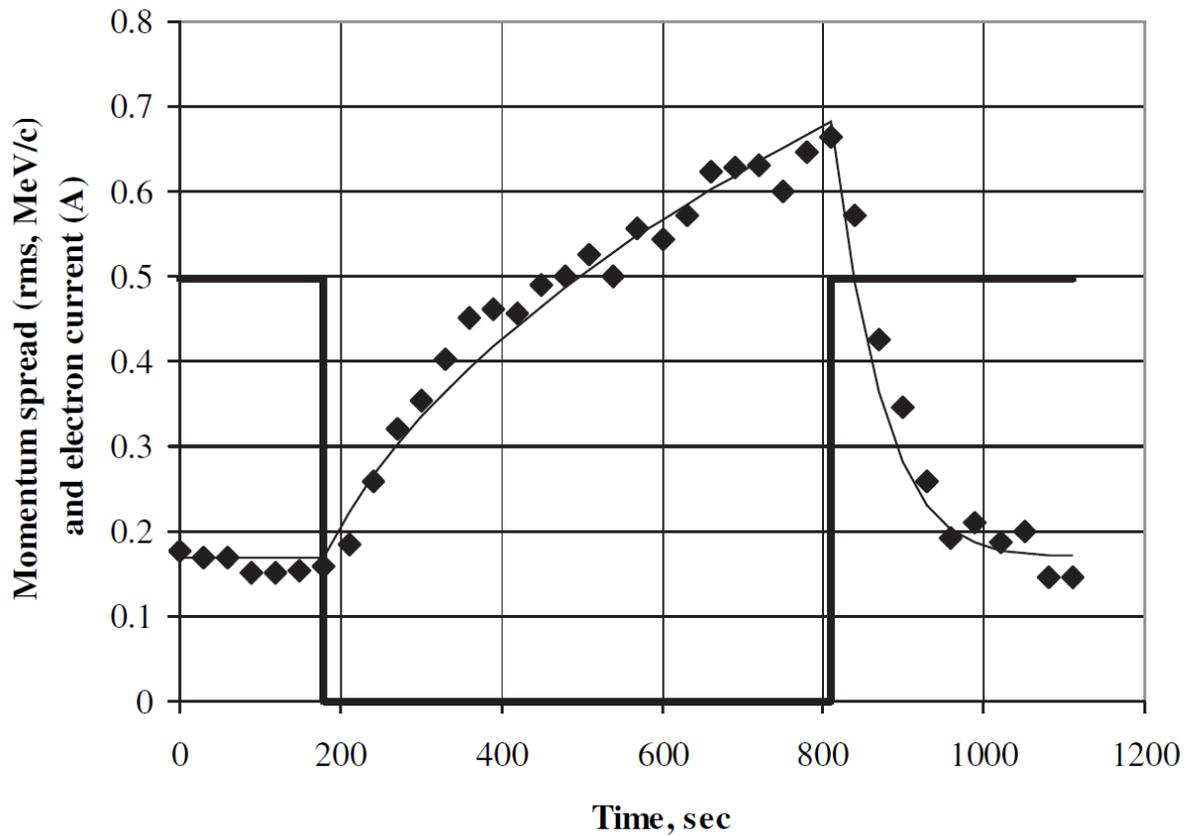

Figure 5: An early demonstration of electron cooling (figure reproduced from (43)). The data points show the rms momentum spread in the antiproton beam, while the bold curve indicates the electron beam current.

.

### 3.3 Beam-beam Effects with Multiple Bunches

The term "beam-beam effects" usually refers to beam losses and emittance growth caused by macroscopic electro-magnetic forces generated by head-on and long-range interactions of colliding beams. The figure of merit for the beam-beam interaction is the induced incoherent tune shift, which is given by:

$$\xi = N_{IP} \frac{N_p r_p}{4\pi\varepsilon_n}$$

(4)

where $r_p = 1.53 \times 10^{-18}$ m denotes the classical proton radius, $N_p$ and $\varepsilon_n$ are the opposite bunch intensity and the rms emittance respectively, and $N_{IP}$ is the number of encounters (44,45). With 36 proton and 36 antiproton bunches the number of encounters per revolution was 72. Through the utilization of electrostatic separators all 72 of these encounters were "long-range" during injection and acceleration, and 70 were long-range during a store – with two head-on encounters remaining at the CDF and D0 detectors. The Tevatron was able to operate consistently at a value of $\xi \approx 0.025 - 0.030$. The vertical and horizontal tunes were set above the half integer between the 5th and 7th order resonances (between 3/5=0.6 and 4/7=0.571) and the beam-beam tune spread fully covered the available tune area – see Figure 6.

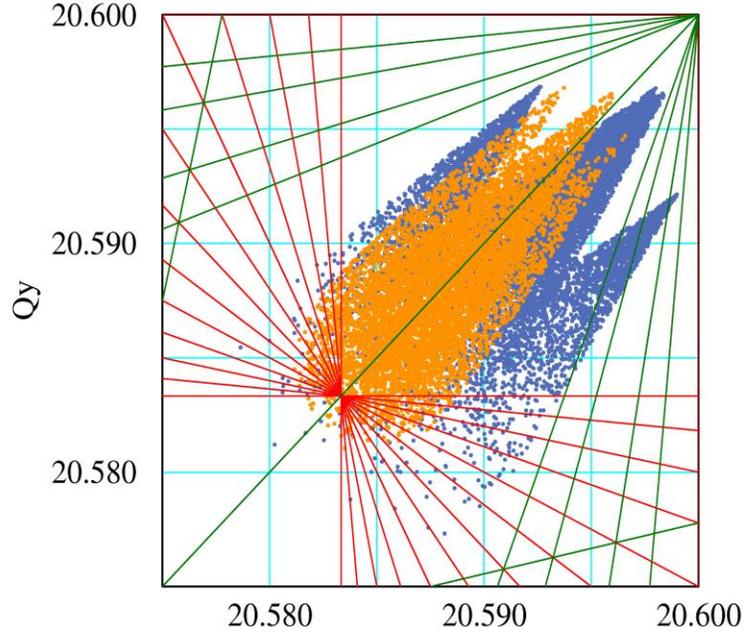

Figure 6: Tevatron proton and antiproton tune distributions in collisions superimposed onto a resonance line plot. The red and green lines are various sum and difference tune resonances of up to 12th order. The blue dots represent calculated the tune distributions for all 36 antiproton bunches; the yellow represent the protons. The tune spread for each bunch is calculated for particles up to 6σ amplitude taking into account the measured intensities and emittances.

The utilization of electrostatic separators to create helical orbits was critical to reducing detrimental beam-beam effects at all stages of injection, acceleration, during the low-beta squeeze, and in collision. Because of the arrangement of proton and antiprotons into three bunch trains the strength of the beam-beam interactions varied from bunch to bunch – in particular the leading and trailing bunches in a train (as evident in Figure 6). Over the course of the Collider Run II several helix designs were employed operationally with a goal of maximizing the *minimum* value of the so-called *radial separation, S,* over all possible parasitic interaction crossing points in units of the RMS betatron beam sizes $\sigma_{x,y\beta}$: $S = \sqrt{(\Delta x / \sigma_{x\beta})^2 + (\Delta y / \sigma_{y\beta})^2}$ (here $\Delta x$ and $\Delta y$ are the distances between the proton and antiproton beam centers). Experience showed that radial separations smaller than $S\sim5$ lead to high beam losses, and it was possible to avoid such separations over all the stages except for a few steps within the low-beta squeeze. The maximum helix amplitude was limited by aperture restrictions at the injection energy and at the beginning of acceleration, and by the maximum sustainable separator gradient, with no sparks, of 48 kV/cm at higher energies. "Feed-down circuits" were used for separate control of focusing lattice parameters on the proton and antiproton orbits: eight sextupole magnet families allowed differential tune and coupling corrections (at injection and at collisions), and four octupole magnet families provided control of differential chromaticities.

During Collider Run II beam losses during the injection, acceleration, and squeeze phases were mostly caused by the long-range beam-beam effects. The beam intensities decayed with the time approximately as $N(t) = N_0 e^{-\sqrt{t/\tau}}$ , indicative of particle diffusion onto physical or dynamic apertures due to the intrabeam scattering (IBS), rf system noise, and scattering on the residual gas. The characteristic decay time $\tau$ was inversely proportional to the opposite beam intensity, beam emittance, and chromaticity. Early in Run II the combined losses from injection to the start of the store were as high as 20-25% for each beam. Besides the helix optimizations, several other measures paved the way to reduction of these losses and higher luminosities including: minimization of the operational chromaticity values; reduction of the injected beam emittances; comprehensive realignment of many Tevatron elements to maximize apertures; and smaller longitudinal emittances due to improvements in Main Injector bunch coalescing. As the result, at the end of Run II, the prior-to-store long-range beam-beam induced losses were cut to 8-10% for protons and 4-5% for antiprotons. Due to significantly smaller emittances, antiproton losses were smaller than the proton losses (despite of 3-5 times higher proton intensity).

The evolution of luminosity during a store was well parameterized as (44):

$$L = \frac{L_0}{\left(1 + \frac{t}{\tau_L}\right)} \qquad (5)$$

Contributions to $\tau_L$ come from both beam loss, primarily driven be beam-beam effects, and emittance growth, primarily driven by IBS. After initiation of collisions beam intensities decayed due to a complex interplay of long-range and head-on beam-beam interactions which led early in Run II to a corresponding loss of the integrated luminosity of about 30-40%. At the end of Run II operations, both species had about the same beam-beam tune shifts because of much smaller antiproton emittances available after 2005 due to the electron cooling of antiprotons in the Recycler. Therefore, the antiprotons effectively experienced only the linear part of the head-on beam-beam force, with rather benign results. Protons on the other hand had tunes closer to 12[th] order resonances and were larger transversely than the anti-protons. Consequently, during head-on collisions, they experienced non-linearities in the beam-beam force enhanced by chromatic effects, and had higher beam loss rate and emittance growth. Increase of radial separation at the first long-range beam crossing outside the interaction points, continuous tune control and stabilization, smaller emittance beams, operation with minimal chromaticity of about +(2-5) units and, especially, correction of originally large chromatic variation of $\beta$* - from $(d\beta*/\beta*)/(dP/P) \approx 500$ down to less than 50 – together resulted in significant reduction of the beam loss rates during the final years of the Collider Run II: to 1.5-2.0%/hour for protons and 1.0-1.5%/hour for antiprotons early in stores.

With beam-beam effects minimized, the total luminosity decay rate of about 20%/hour was mostly determined by emittance growth and by particle burn-up due to collisions. The integrated luminosity per store depends on the product of the initial luminosity and the luminosity lifetime $\tau_L$: $\int L dt \approx L_0 \tau_L \ln(1+T/\tau_L)$. Therefore, the combined impact of all the beam-beam effects (before the collisions and during the stores) on the luminosity integral was ~(20-30)%.

The beam-beam effects in the Tevatron caused all measurable indicators of beam dynamics, not just the tunes, to vary as a function of position within a bunch train. For example, the helical orbits of antiproton bunches during the collisions varied along the bunch trains by some 40 to 50 $\mu$m in a systematic, ladder-like fashion; the antiproton tunes varied by 0.005-0.008 from bunch to bunch; and chromaticity varied by up to 6 units – all in in acceptable agreement with theory (46). Variations in the corresponding proton parameters were factor of 3-5 smaller. Consequently, the antiproton and proton bunch intensity lifetimes and, sometimes, the emittance growth rates varied considerably from bunch to bunch. For example, bunches at the ends of each bunch train usually lost intensity significantly faster than other bunches because their vertical tunes were closest to the 7/12 resonance – see Figure 7.

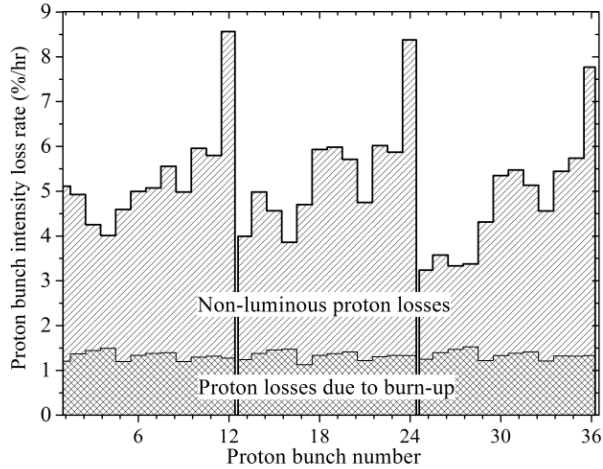

Figure 7: Proton-bunch intensity loss rates $(dN_p/N_p)/dt$ at the beginning of a Tevatron store with an initial luminosity $L$=250×$10^{30}$ cm$^{-2}$s$^{-1}$ (47). The losses due to the burn-up (inelastic proton-antiproton interactions) $dN_p/dt = \sigma_{int} L$ at the two main IPs ($\sigma_{int}$ =0.07 barn) are separated.

Electron lenses were originally developed for compensation of electromagnetic long-range and head-on beam-beam interactions of proton and antiproton beams. Strong space-charge forces from a low-energy high-current electron beams can be precisely controlled and adjusted between individual bunches, equalizing the bunch-to-bunch differences and optimizing the performance of all bunches in a multibunch collider. Two electron lenses were built and installed in the Tevatron. When applied to proton bunches #12, 24, 36, they reduced the proton losses by more than a factor of two (47,48).

### 3.4 Intrabeam Scattering

Intrabeam scattering refers to the effect of small-angle multiple Coulomb collisions between particles in the beam that results in  beam emittance growth in all three dimensions (49). The characteristic IBS growth times for the proton and antiproton bunches within the accelerator complex ranged from several to several dozen hours, significantly impacting the beams circulating for many hours in the Antiproton Accumulator, Recycler Ring and Tevatron. The emittance evolution equation in the presence of cooling is (both transverse and longitudinal):

$$d\varepsilon_{T,L}/dt = (d\varepsilon_{T,L}/dt)_{heating} - \varepsilon_{T,L}/\tau_{T,L\ cooling} \qquad (6)$$

As the dominant heating mechanism for high intensity beams, IBS determined the equilibrium emittances achievable in the Accumulator and Recycler, and defined the luminosity evolution in the Tevatron.

In most common situations, where the longitudinal velocity spread in the beam frame is much smaller than the transverse spread, energy changes due to the individual particles collisions lead to excitation of the betatron motion and predominantly transverse emittance growth with the rate (50,51):

$$\frac{d\varepsilon_{n,rms}}{dt} \approx \frac{r_p^2 Nc}{4\sqrt{2}\gamma^2\beta^2\sigma_z}\left\langle\frac{L_C}{\sigma_x\sigma_y\theta_T}A_x\kappa\right\rangle \qquad (7)$$

where $r_p$ and $c$ are the proton classical radius and the speed of light, $\gamma$ and $\beta$ are the relativistic factors, $N$ is the number of particles, $\sigma_z$ is the rms bunch length, $\sigma_x$, $\sigma_y$, $\theta_T$ are the rms sizes and the rms transverse angular spread, $\kappa$ is the $x$-$y$ coupling parameter (typically 0.3-0.7), $L_C$ is the Coulomb logarithm ($L_C \approx 20$), brackets $<...>$  denote averaging over the ring, and the coefficient $A_x=(D_x{}^2+(D_x{}'\beta_x+\alpha_x D_x)^2/\,\beta_x\,)$ combines focusing lattice Twiss parameters - $\beta_x$ and $\alpha_x$ are the horizontal beta- and alpha-functions, and $D_x$ and $D_x{}'$ are the dispersion and its derivative.

IBS became an important concern within the Accumulator as the operating frequencies of the stochastic cooling systems increased, necessitating a lower phase-slip factor (the proportionality between revolution frequency and momentum) to avoid longitudinal cooling instabilities. This change, introduced in the early 2000's, initially resulted in poor performance due to a significant increase in the beam emittance of the collected antiprotons due to IBS (50). The solution was implementation of a "shot lattice" (52) – the optics of the Accumulator were temporarily modified during a shot to provide a lower emittance prior to the transfer of beam to the Tevatron. This adjustment changed the maximum value of $A_x$ from about 25 m, as needed in the regime of fast antiproton cooling, to about 10 m in the final 30 minutes before beginning of injection to the Tevatron, thus, resulting in about 2.5 times smaller equilibrium emittance of antiprotons.

Within the high intensity antiproton beams stored in the Recycler (typically $(3\text{-}6)\times10^{12}$), the IBS contribution to the overall transverse emittance growth rate of ~0.3 $\pi$ mm-mrad/hr was about twice that from beam-gas scattering. The interplay of strong longitudinal electron cooling, with $\tau_{L\,cooling}<0.5$ hour, and IBS resulted in fast overall 6-dimensional cooling and equilibrium beam phase space distributions with approximately equal transverse and longitudinal velocity spreads in the beam frame. Beam studies showed that, depending on the initial rms momentum spread, the longitudinal IBS heating could vanish or even become negative (i.e., the IBS lead to cooling in one degree of freedom at the expense of another one) (53).

At the end of Collider Run II, with the Tevatron achieving initial luminosities in the range of 3-4×10$^{32}$ cm$^{-2}$ s$^{-1}$, the luminosity lifetime was about $\tau_L$=-$L/(dL/dt)$=5.2-5.7 hours with the largest contribution coming from transverse beam emittance growth with a characteristic time of $\tau_\varepsilon$=$\varepsilon_T/(d\varepsilon_T/dt)$ of about 10-13 hours. This growth was dominated by IBS in the proton and antiprotons bunches, with small contributions from the beam-gas scattering and external noise sources. Note that for long stores the integrated luminosity scales approximately as the product of initial luminosity and the luminosity lifetime $\tau_L$, therefore, the intrabeam scattering accounted for some 40%-50% reduction in the overall luminosity integral.

Several beam experiments were conducted to separate contributions of different phenomena to the emittance growth. In one (54), 15 proton bunches with various intensities were accelerated to 980 GeV and underwent the low-beta squeeze. The bunches had variable transverse emittances $\varepsilon_T$ in the range from 2.3 to 3.6 $\pi$ mm-mrad, and rms bunch lengths $\sigma_z$ in the range 1.7 to 2.1 ns. The bunches were stored in the Tevatron for 3 hours with their emittances, bunch lengths and intensities continuously monitored. The IBS-driven emittance growth should be proportional to the factor $F_{IBS} = \dfrac{N_p}{\varepsilon_T^{1.5}\sigma_z}$ . For the nominal high luminosity operation regime that factor reaches ~25-30 [10$^9$/ mm mrad $^{3/2}$ /ns] in both proton and antiproton bunches. The observed emittance growth (Figure 8) and the rms bunch length indeed grew with $F_{IBS}$ approximately as:

$$\frac{d\varepsilon_V}{dt}[\pi\, mmmrad/hr] = (0.072 \pm 0.02) + (0.0079 \pm 0.0015)\cdot F_{IBS} \qquad (8)$$

$$\frac{d\sigma_z^2}{dt}[ns^2/hr] = (0.0063 \pm 0.0193) + (0.0173 \pm 0.0013)\cdot F_{IBS} \qquad (9).$$

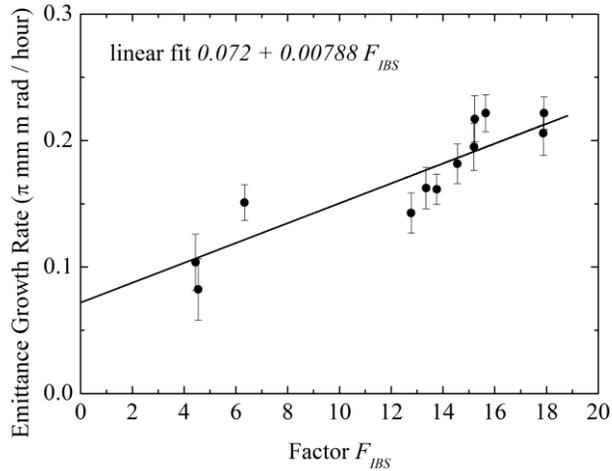

Figure 8: Vertical emittance growth rates of proton bunches vs the IBS factor $F_{IBS}$ (see text).

The intercept (zero intensity) transverse emittance growth of 0.072 $\pi$ mm mrad/hr is due to the intensity independent effects such as external transverse noise (~80%) and scattering on the residual vacuum molecules (~20%) (55). The longitudinal diffusion is totally dominated by IBS, and the tiny zero intensity bunch lengthening is due to small rf phase noise caused by microphonics effects in the rf cavities excited by the flow of cooling water.

### 3.5 RF Manipulations

A number of rf manipulations were critical in the high luminosity operation of the Tevatron Collider. These included bunch rotations, bunch coalescing, bunch cogging, slip-stacking, and a variety of barrier bucket manipulations. The first three of these were essential features dating from initial operations in the 1980's (56, 57). Very briefly, these manipulations were used to shorten the proton bunches impinging on the antiproton production target, to consolidate several modest intensity bunches into a single high intensity bunch, and to adjust the relative (azimuthal) positions of protons and antiprotons within the Tevatron. More novel developments that occurred later during the Tevatron era were the implementation of slip-stacking in the Main Injector and the utilization of barrier buckets to manipulate the antiproton beam in the Recycler.

### 3.5.1 Slip-stacking

Slip-stacking is a longitudinal stacking technique first demonstrated at CERN (58). The technique was extended to high intensities in the Main Injector in order to double the number of protons delivered onto the antiproton production target (59). The slip-stacking mechanism relies on the correlation between momentum and revolution frequency in a synchrotron – bunches separated around the circumference of the ring can be aligned and merged if they are maintained at different momenta. In its simplest form a bunch train occupying $1/7^{th}$ of the circumference of the MI was injected from the Booster. The 18 rf cavities in the MI rf system were divided into two groups operating at different frequencies. The first injected bunch train was captured on the central orbit by the subset of the rf cavities operating at the nominal rf frequency (about 52.8 MHz). This bunch train was then accelerated by about 30 MeV (1400 kHz change in the rf frequency). A second bunch train was then injected immediately adjacent to the first train, and captured at the nominal rf frequency. The higher energy bunches would then "catch up" to the lower energy bunch in a few milliseconds, at which point all rf cavities were tuned to a common frequency and the two sets of bunches were merged into a single set of rf buckets and subsequently accelerated to 120 GeV.

In practice this process was carried out multiple time in order to fit 11 bunch trains from the Booster into an extended train occupying 6/7$^{th}$ of the MI circumference. This allowed operations for antiproton production and for neutrino production to proceed simultaneously: one-sixth of the extended bunch train was directed onto the antiproton target and the remainder onto the neutrino target. The process was quite efficient with roughly 95% of the injected protons being stacked, accelerated, and delivered onto one of the two targets. The high efficiency was a direct result of the generous momentum aperture of the Main Injector and an advanced beam loading compensation system. The net result of slip-stacking was a factor of two increase in protons delivered to the antiproton target. Figure 9 shows the performance of the system in practice.

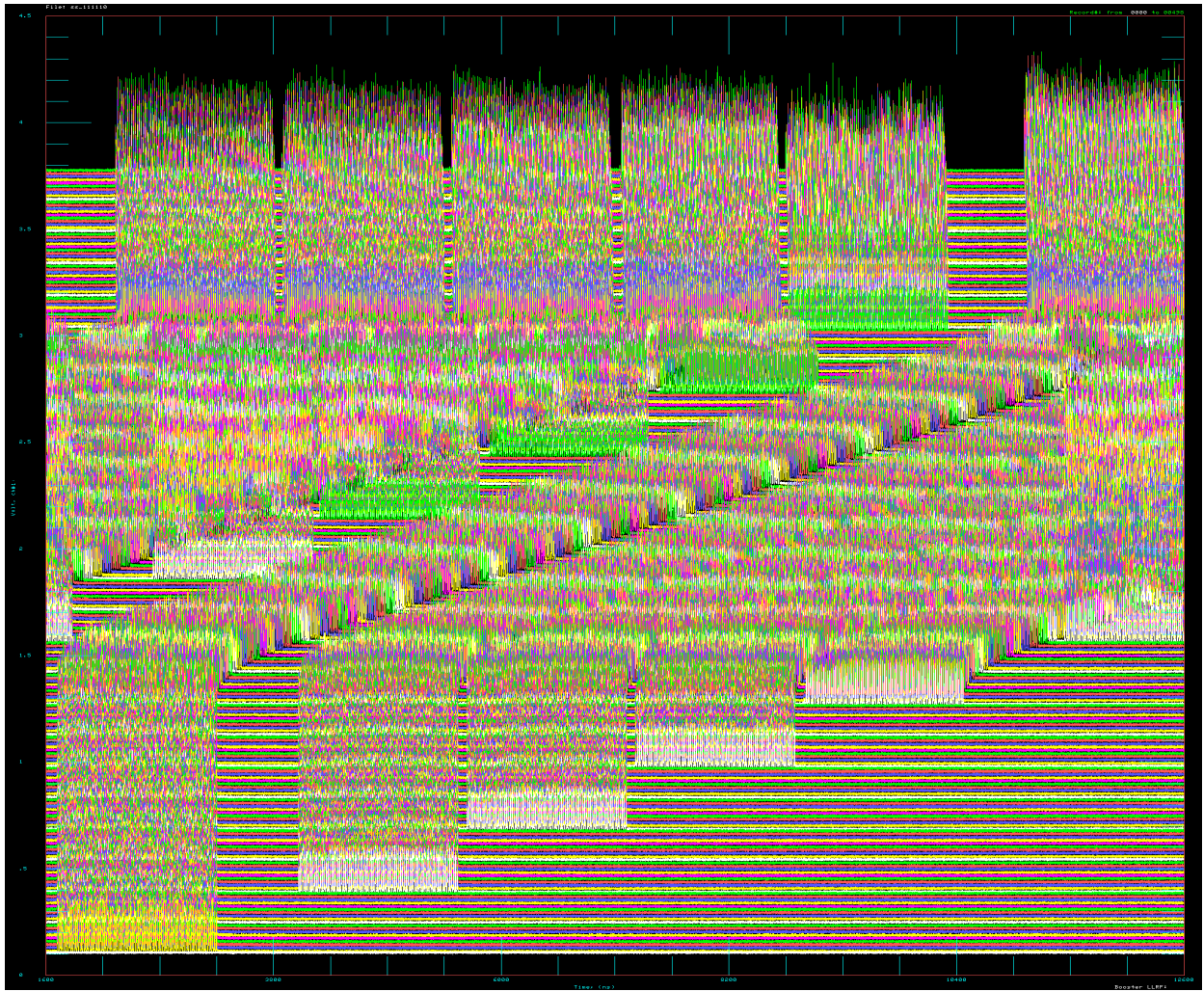

Figure 9: Slip-stacking in the Main Injector. Shown are the injection of 11 bunch trains and their subsequent merging into six trains (time proceeds from bottom to top). The beam intensity signal is shown relative to the circumference of the Main Injector (3319 m, 11 μsec) on the horizontal axis, and versus time on the vertical axis (roughly 750 msec full scale).

3.5.2 Barrier Bucket Manipulations

The term "barrier bucket" refers generally to a voltage waveform that is unipolar and limited in (temporal) extent, rather than the more commonly utilized continuous sine wave. Examples include both half sine wave and rectangular waveforms. Such waveforms present a barrier to beam particles that are not traveling with the synchronous momentum, and hence the beam in a circular accelerator can be confined (or excluded) between two such waveforms of opposite polarity. The applications of barrier buckets include longitudinal confinement of the beam with a uniform density (an advantage for beam cooling), segregation (around the circumference of the machine) of beam segments with differing characteristics, and injection/accumulation of beams.

The first example at Fermilab was the utilization of a simple barrier system consisting of a single cycle of a sine wave with a time duration one-quarter the revolution period of the Debuncher ring. This system provided a gap within the otherwise continuous antiproton beam to accommodate the risetime of the extraction kicker and the smaller circumference of the Accumulator.

Within the Recycler more sophisticated barrier bucket capabilities formed an essential element of both the antiproton accumulation process and the preparation of antiprotons for transfer to the Tevatron (60, 61). The Recycler rf system was designed to allow the generation of multiple voltage pulses of differing shapes, amplitudes, and lengths, and with variable distributions around the ring. The system was extremely flexible in enabling complex manipulations of beams in longitudinal phase space. The drive signals were developed through digital signal processors driving arbitrary waveform generators into broadband (10 kHz to 100 MHz) amplifiers. Four ferrite loaded cavities were capable of generating a total of 2-3 kV. A linearization system was required to create a uniform beam density in the presence of distortions in the rf system, beamloading, and potential well distortions.

Antiproton accumulation in the Recycler proceeded through transfers of roughly $1 \times 10^{11}$ antiprotons from the Accumulator every 30 minutes. These transfers were implemented by opening up a barrier bucket that would exclude the already accumulated antiprotons from a portion of the circumference of the Recycler corresponding to the length of the injected antiproton pulse. Once the new antiprotons were injected the barrier buckets were removed and the ensemble of new and old antiprotons was cooled prior to the next injection.

A more sophisticated operation was implemented upon extraction of antiprotons from the Recycler. This manipulation, designated "momentum mining", took advantage of the characteristics of electron cooling to extract the highest (longitudinal) density antiprotons for delivery to the Tevatron. Momentum mining provided the 36 antiproton bunches required for the Tevatron with very uniform populations, and minimal transverse and longitudinal emittances. The momentum mining procedure is displayed in Figure 10 with three sets of traces showing the mining steps. In each set the upper signal is the rf waveform and the lower trace is the beam signal. The shaded area represents one Recycler circumference. Trace a) shows the longitudinal profile of antiprotons before mining is initiated – the antiprotons are confined to 6.1 μsec of the 11.1 μsec Recycler circumference. Trace b) shows the transformation of these antiprotons into nine distinct bunches. The barrier buckets have a momentum acceptance that is less than the full momentum spread in the beam, hence the excluded high energy tail. In trace c) one of the nine bunches is segregated and further split into four bunches with the 396 nsec spacing desired in the Tevatron. The four bunches are then extracted and sent to the Tevatron via the Main Injector. The step represented by trace c) is repeated nine times in total to provide 36 antiproton bunches to the Tevatron. The remaining high energy tail is retained as the starting point of the subsequent accumulation process. The result was a highly efficient utilization of antiprotons.

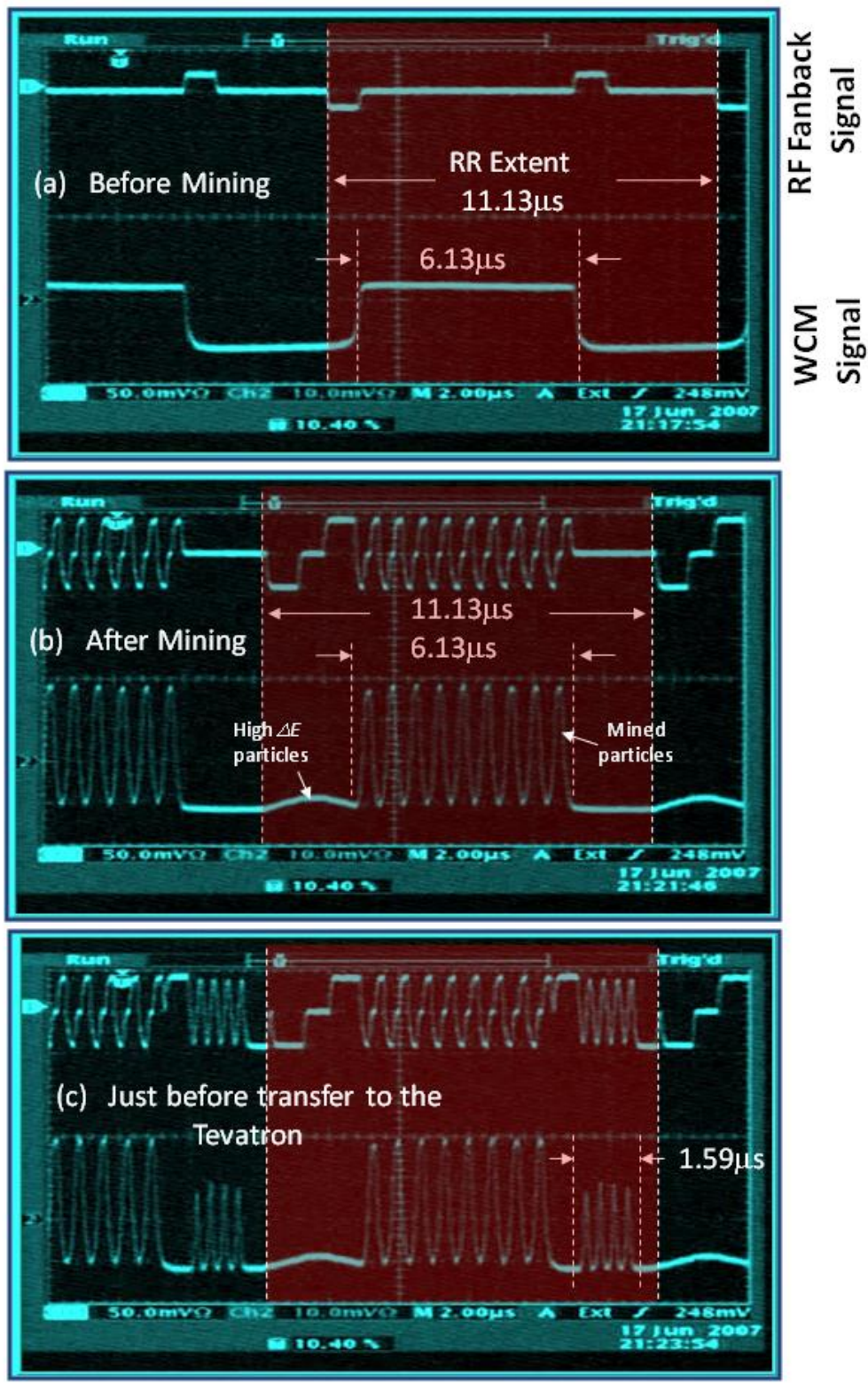

Figure 10: Momentum mining in the Recycler. (See text for explanation.)

### 3.6 Beam Instabilities

High luminosity operation of the Tevatron during Collider Run II required high beam intensities and, as the result, all the rings (except the Debuncher) had notable problems with beam stability. Instabilities of almost every type were present – single and multi-bunch, transverse and longitudinal – due to electromagnetic interactions with the vacuum chamber and with ions captured in the beams. Several methods to suppress the instabilities were implemented, including various damping systems.

The vacuum chamber of the Booster synchrotron is formed by the laminated poles of the combined function magnets. The magnet laminations contribute a total transverse impedance of about $Z_\perp \approx$40-55 M$\Omega$/m and affect beam stability (62). In operation, coherent synchro-betatron instabilities were suppressed by large chromaticities $Q'\sim$(10-16) (63), which lead to deterioration of the dynamic aperture and beam lifetime reduction. Attempts to stabilize the beam by transverse feedback systems were not fully successful. At chromaticities below the threshold, multibunch high order head-tail modes developed with growth time of 12-18 turns (64). The operational configuration of the Booster was established as a compromise between these effects.

In the Recycler electron cooling led to small emittances and a fast (< 0.1s) transverse resistive wall instability occurring beyond a certain threshold in 6-dimensional beam brightness, $N_a/(\varepsilon_x\,\varepsilon_y$ $\varepsilon_L$). A transverse damper system was installed in 2005 with an initial bandwidth of 30 MHz, and eventually upgraded to 70 MHz and allowed stable operation with very bright antiproton beams (65).

The transverse "weak head-tail instability" was a serious limitation on the maximum proton current at the Tevatron injection energy of 150 GeV (66). It manifested itself as a very fast (50-100 turns) development of vertical or horizontal oscillations, and consequent beam loss accompanied by simultaneous emittance blowup of many bunches in a bunch train. For a long time, the only way to stabilize it was to operate the Tevatron at high chromaticity (>10) in both planes, leading to very short beam intensity lifetime, especially, in the presence of the opposite beam. A combination of transverse bunch-by-bunch dampers, installation of 0.4 mm thin conductive CuBe liners inside the injection (Lambertson) magnets that reduced the total Tevatron transverse impedance from $Z_\perp \approx$ (2.4-5) M$\Omega$/m to about 1 M$\Omega$/m, and operation with octupole magnets to increase the tune-spread for Landau damping eventually resulted in overall stability and better beam lifetime at much lower chromaticities of about 0-3.

Longitudinal single bunch and coupled bunch instabilities occurred in the Tevatron – e.g., at the Tevatron injection energy of 150 GeV very large (up to $60^\circ$) proton bunch rf phase oscillations could persist for many minutes (67). Large beam oscillations also regularly occurred at 980 GeV and resulted in significant longitudinal emittance increase, reduction of luminosity, proton beam losses, and accumulation of particles in the abort gaps. In the later years of the Collider Run II, similar phenomena started to appear irregularly in the high intensity antiproton beams. To counteract these phenomena, a longitudinal bunch-by-bunch damper was designed, built, installed, and commissioned (only for protons) in the Tevatron (68). It effectively suppressed both the "dancing bunches" phenomena and the single and coupled bunch instabilities. It was found that to be effective, the damper gain should vary slowly during the store in a fashion which tracks the proton longitudinal density $N_p/\sigma_z$.

## 3.7 Detector Backgrounds

Even under good operational conditions a finite fraction of the beam will leave the stable central area of an accelerator because of beam-gas interactions, intra-beam scattering, proton-antiproton interactions, rf noise, ground motion, and resonances excited by the accelerator elements imperfection. These particles form a beam halo. As a result of halo interactions with limiting apertures, hadronic and electromagnetic showers are induced in accelerator and detector components causing numerous deleterious effects ranging from minor to severe. The most critical for colliders are accelerator related backgrounds in the collider detectors and beam losses in superconducting magnets.

During Collider Run I the Tevatron halo removal system experienced limitations that prompted design of a new system (69) for Run II, which was built to localize most of the losses in the straight sections away from the CDF and D0 detectors. The system incorporated four (vertical and horizontal, proton and antiproton) primary collimators/targets (5 mm W targets set some $5\sigma$ from the beam axis) and eight secondary collimators (1.5 m stainless steel set about $6\sigma$ away from the beam axis). This highly automated system allowed entire halo removal process to take place in approximately 7 min.

The halo removal process was initiated by the Tevatron sequencer at 980 GeV after the proton and antiproton beams had been brought into collisions. There were 4 sub-sequence operations that were necessary in order to complete halo removal:  1) *Move Collimators to Initial Positions*: Move all collimators at 1.25mm/sec to the "half way" point to the beam; 2) *Intermediate Halo Removal*: Each set (proton and antiproton) of collimators and targets were moved together under beam loss monitor feedback until a small loss was detected and all collimators in the set stopped; 3) *Perform Halo Removal*: Each secondary collimator and target moved serially into the beam. Secondary collimators were moved under loss monitor feedback with a step size of  0.025mm until they reach the edge of the beam to shadow the losses by the primary collimator.  After all secondary collimators were placed next to the beam, each target was moved under loss monitor and beam intensity feedback until 0.4% of each beam (proton and antiproton) was removed; 4) *Retract Collimators For Store*: After targets and secondary collimators have reached their final positions, they were retracted approximately 1mm.  This is the position they remained in for the duration of the store.   Figure 11 shows the detector background rates during the process of beam collimation early in a store.

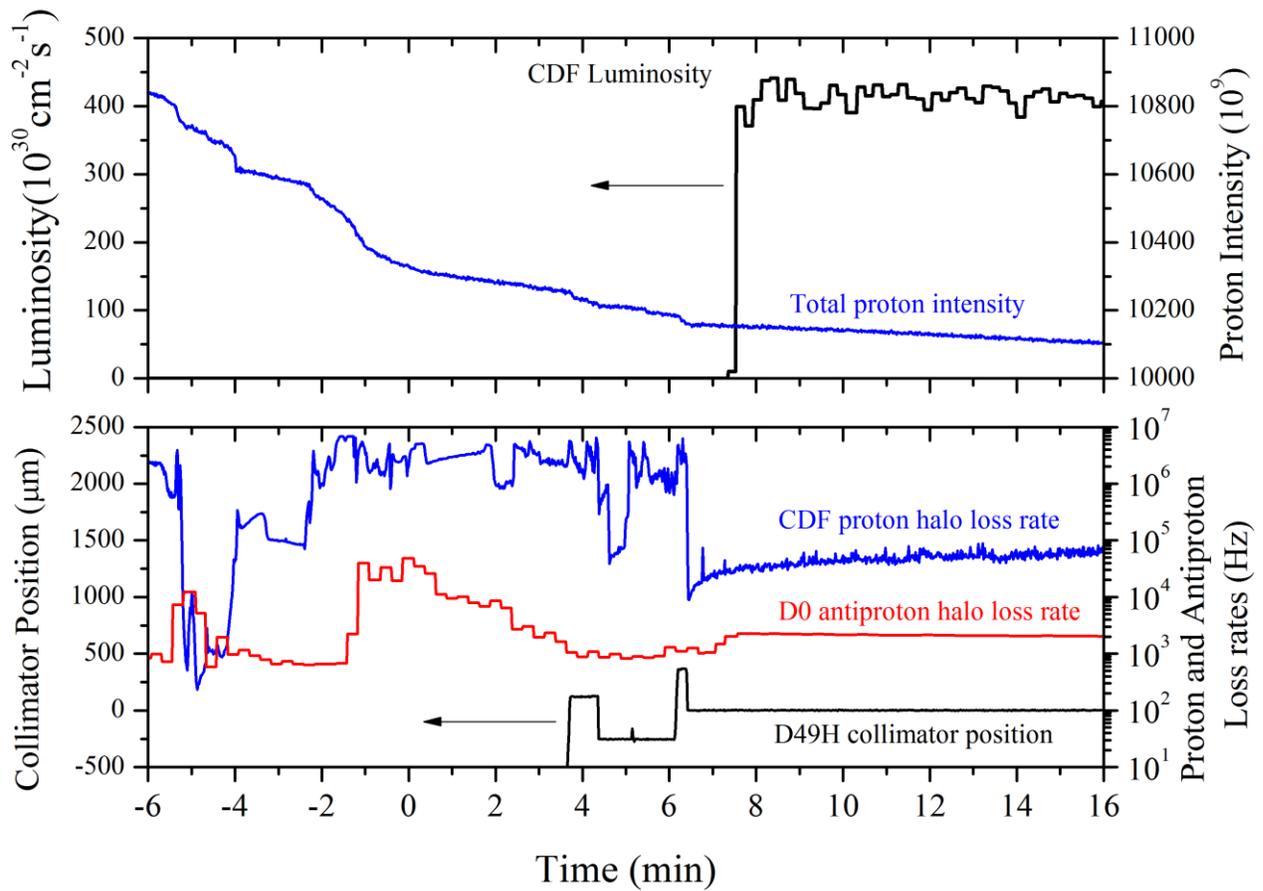

Figure 11: Collimation process early in a store. Zero time corresponds to the moment when the proton and antiproton beams are brought to collisions. The upper plot shows proton beam intensity and CDF luminosity. The bottom plot shows beam halo loss rates as measured by CDF and D0, and the position of one of the collimators (D49H).

The system was upgraded several times following operational needs. For example, in 2002, the Tevatron Electron Lenses (TEL) was set up to remove uncaptured (by the rf system) particles from the abort beam gaps and, thus, to reduce the risk of beam induced damage to CDF and D0 during beam aborts (70). The high current, low energy, electron beams of the TELs were set in close proximity to proton or antiproton orbits and generated strong transverse kicks of about 0.07 μrad (for 5kV electrons with typical peak current of about 0.6 A and 5 mm away from the protons). As the TELs possessed short rise and fall times (~100 ns), they were timed into all three 2.6 µs gaps between bunch trains and pulsed every $7^{th}$ turn to excite resonantly betatron oscillations of unwanted particles in the gaps as the tune of the machine was close to the 4/7 resonance. These particles were removed by the collimators, thereby preventing the accumulation of the uncaptured beam in the abort gaps.

In 2003, following several instances of unsynchronized abort kicker pre-fires in the Tevatron, an additional *tertiary* collimator was installed at the A48 location to protect the CDF detector. Beam scraping procedures optimized for faster operation and highly-efficient repeated scraping (double scraping) of the beams were made operational in 2005. In 2010, collimation during a portion of the low-beta squeeze was added in order to reduce losses at CDF and D0 that were causing frequent quenches. One collimator was placed at $5\sigma$ to become a limiting aperture and redirect the losses away from CDF and D0 to a region that had more robust magnet quench limits.

The halo removal efficiency, defined as the ratio of the halo losses at the detectors before and after the halo removal, was about 110 and 80 for the CDF proton and antiproton losses, respectively, and 13 and 20 for the D0 proton and antiproton losses. The reduction in the D0 proton halo loss was relatively small because the CDF interaction region provided additional collimation of the proton halo losses upstream of D0.

**3.8 Future Techniques**

Two novel ideas to improve beam collimation efficiency were extensively studied during Collider Run II – bent crystals and hollow electron beams. A bent crystal can coherently channel halo particles deeper into a nearby secondary absorber and, therefore, can provide a substitute for the primary target and improve beam collimation efficiency. Such high efficiency collimation by bent Si crystals was successfully demonstrated at the Tevatron (71, 72). Studies with a variety of crystal configurations identified and studied experimentally several processes that take place during the passage of protons through the crystals: a) amorphous scattering of the primary beam; b) channeling – by as much as 0.1-0.4 mrad; c) dechanneling due to scattering in the bulk of the crystal; d) "volume reflection" off the bent planes by as much as 0.05-0.2 mrad; and e) "volume capture" of initially unchanneled particles into the channeling regime after scattering inside the crystal (73).

The hollow electron beam collimator is a new concept for controlled halo removal of intense high-energy hadron beams in storage rings and colliders (74). It is based on the interaction of the circulating beam with a 5 keV, magnetically confined, pulsed hollow electron beam produced by one of the TELs. The electrons enclose the circulating beam, kicking halo particles transversely and leaving the beam core unperturbed (Figure 12). By acting as a tunable diffusion enhancer rather than as a hard aperture limitation, the hollow electron beam collimator extends conventional collimation systems beyond the intensity limits imposed by tolerable losses. The concept was successfully tested experimentally in the Tevatron (75).  This technique represents a promising option for scraping high-power beams in the LHC.

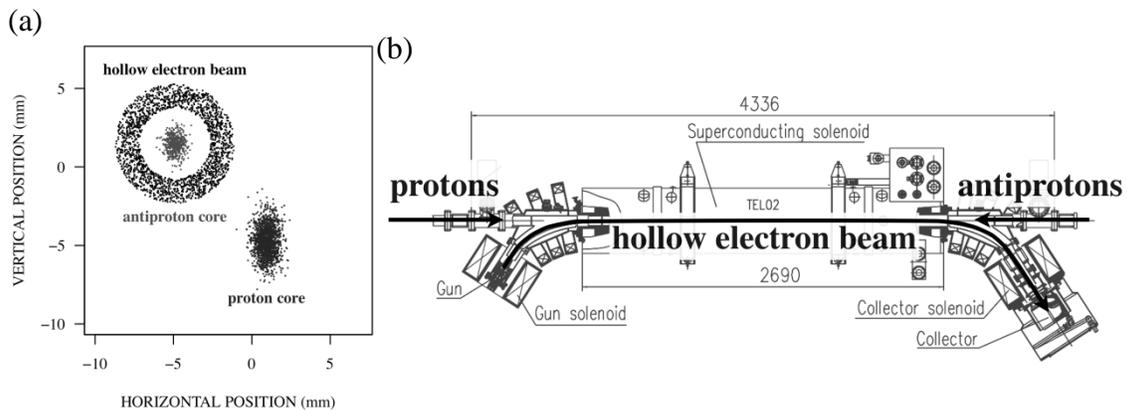

Figure 12: Hollow electron beam collimator: (a) transverse beam layout; (b) top view of the beams in the Tevatron Electron Lens.

The need for precise control of the Tevatron betatron frequencies led to development of several innovative tune measurement systems (76). Multi-GHz Schottky monitors were successfully employed for multi-bunch non-invasive diagnostics of several beam parameters in the Tevatron, Recycler and, later, in the LHC (77). The Tevatron 1.7 GHz Schottky pickups were slotted waveguide structures with bandwidths exceeding several hundred MHz. Since the devices were not resonant, they were gated on select bunches, making possible bunch-by-bunch tune measurements for protons and antiprotons with accuracy of about 0.001 – see Figure 13. The 1.7 GHz tune readings were used in everyday operation to stabilize the antiproton tunes as the beam-beam tune shift changed over the course of a store. Chromaticity, momentum spread and emittance were also extracted from the signals, making the 1.7 GHz Schottky monitors a very versatile tool.

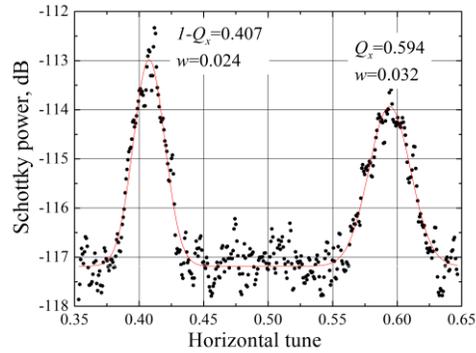

Figure 13: The Tevatron horizontal proton Schottky spectra around 1.7 GHz, showing both upper and lower betatron sidebands.

Long- and short-term stability of the beam orbits was of utmost operational importance for the Tevatron. The beam orbits wandered during multi-hour long stores by as much as 0.1-0.4 mm, and 0.5-1 mm over periods of weeks. Ground motion was found to be the source, and as monitors all Tevatron magnets were equipped with hydrostatic level sensors (HLS), while dozens of fast 1 µrad resolution tiltmeters and 0.1 µm resolution HLSs were set on most critical low-beta quadrupoles near the two IPs (78). Various sources of ground motion such as tidal motion, subsidence caused by earthquakes, thermal drift, pumping of ground water, traffic and quenches of the cryogenic magnets were observed. A geophysical phenomena of ground diffusion governed by the so-called "*ATL law*" (79) has been identified. To keep the orbits stable during stores within 50 µm, an automated orbit-smoothing algorithm was implemented which used several dipole corrector magnets (80).

# 4. SUMMARY OF PERFORMANCE HISTORY AND LEGACY

The initial design luminosity of the Tevatron was $10^{30}$ cm$^{-2}$s$^{-1}$; however after two decades of developing understanding of limitations and implementation of associated upgrades the accelerator was able to deliver 430 times higher luminosities to the CDF and D0 experiments. The Collider performance history (Figure 14) shows that the luminosity increases occurred after numerous improvements, some of which were implemented during operations and others introduced during regular shutdown periods. Improvements took place in all accelerators of the Collider complex and addressed all parameters affecting luminosity – proton and antiproton intensities, emittances, optical functions, losses, reliability and availability (81). The pace of the Tevatron luminosity progress was among the fastest of any high energy collider (82). The Tevatron worked extremely well for 25 years and delivered more than 12 fb$^{-1}$ of the integrated luminosity to each detector (CDF and D0) before being shut off on September 30, 2011. The Tevatron Collider was arguably one of the most complex research instruments ever to reach the operations stage and is widely recognized for numerous physics discoveries and many technological breakthroughs.

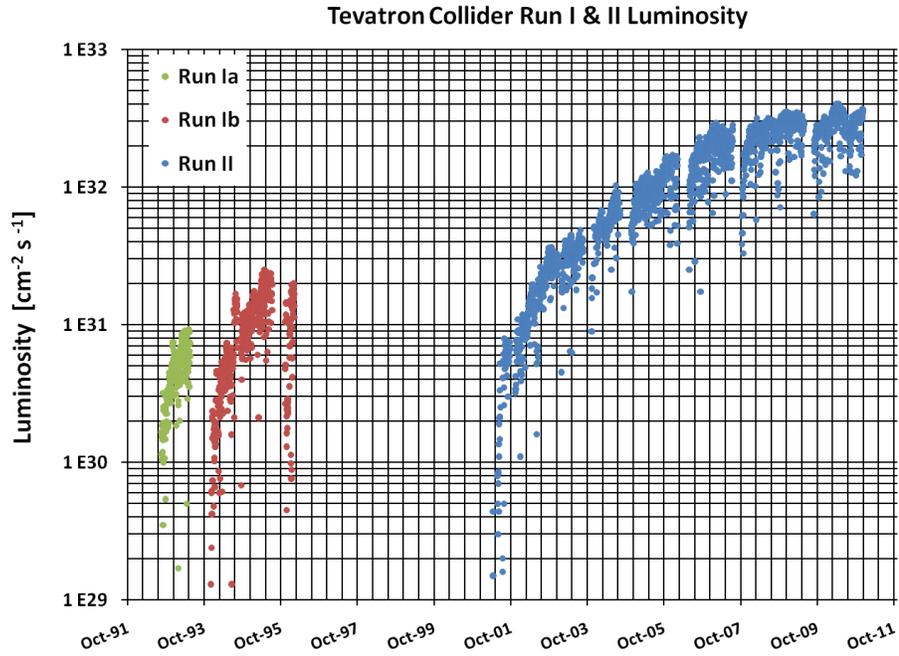

Figure 14: Initial luminosity for all stores in Collider Runs I and II

Operation of such complex systems required persistence and stubbornness, close attention to details, and a willingness to purse incremental improvements rather than search for "silver bullets". Improvements required flexibility in considering all possibilities for increasing luminosity, and a willingness to change plans if/when experience showed prospects diminishing due to the complexity of machines and/or unpredictability of the performance limits. The Tevatron Collider plans changed as soon as difficulties were encountered at the beginning of Run II – reasons were understood and methods developed under the "Run II Luminosity Upgrade" plan to achieve superb performance. Expectations management was also very important. Unavoidable operational difficulties not only generated strain, but also inspired creativity in the entire team of scientists and engineers, managers and technicians, support staff and collaborators.

Among the most notable contributions of the Tevatron to the accelerator technology and beam physics are the development and first large scale deployment of accelerator quality superconducting high field magnets; the first high energy accelerator ever built with permanent magnets; the world's most advanced antiproton production and cooling complex which delivered more than 90% of the world's total man-made nuclear antimatter (17 nanograms) over the decades; the highest performance antiproton stochastic cooling and electron cooling systems; the first extensive operational use of slip-stacking and barrier bucket rf manipulation methods; electron lenses for beam-beam compensation and hollow electron beam collimation; great advancements in beam physics through the studies of the beam-beam effects, crystal collimation, electron cloud (83) and beam emittance growth mechanisms; new theories of beam optics (84,85), intra-beam scattering and instabilities (86); sophisticated beam-beam and luminosity modeling (87,88); and more efficient beam instrumentation (89,90,91,92,93,94,95,96).

The story of the Tevatron remarkably illustrates how the exchange of ideas and methods and technology transfer helps the field of accelerators: Fermilab scientists and engineers learned and borrowed a great deal of knowledge from CERN's ISR and SPS accelerators, and they interacted directly with industry on critical superconducting cable technologies - in the words of the late Robert Marsh of Teledyne Wah Chang, a leading supplier of superconducting alloys (97), "Every program in superconductivity that there is today owes itself in some measure to the fact that Fermilab built the Tevatron and it worked." In turn, the technologies, techniques, and operating experiences at the Tevatron have been successfully applied to the subsequent generation of high energy hadron colliders: HERA (5), RHIC (6), and LHC (7). Finally, the multitude of accelerator achievements and breakthroughs of the Tevatron have laid the foundations for the next generation of accelerators devoted to particle physics research (98) such as Project X (99), the LHC upgrades (100) and the Muon Collider (101).

## ACKNOWLEDGEMENTS

The authors represent literally thousands of people who achieved great success by working together over many years with great skill, enthusiasm and dedication. Scientists, engineers, designers, drafters, programmers, technicians, secretaries, buyers, contract administrators, safety professionals, lawyers, financial administrators, janitors, cafeteria workers – all put their hearts into the Tevatron and are responsible for this historic accomplishment.